# Spin Angular Momentum Transfer and Plasmogalvanic Phenomena

Maxim Durach[1*], Natalia Noginova[2]

1. Georgia Southern University; 2. Norfolk State University

*email: mdurach@georgiasouthern.edu

**Abstract.** We introduce the continuity equation for the electromagnetic spin angular momentum (SAM) in matter and discuss the torque associated with the SAM transfer in terms of effective spin forces acting in a material. In plasmonic metal, these spin forces result in plasmogalvanic phenomenon which is pinning the plasmon-induced electromotive force to atomically-thin layer at the metal interface.

Investigation of angular momentum in electromagnetic fields has been a rapidly developing research direction. The major results are separation of light angular momentum into orbital angular momentum (OAM) and spin angular momentum (SAM) parts [1,2], conservation laws for angular momentum [3] and OAM and SAM separately [4], and spin-momentum locking in evanescent waves [5-7]. Recently the subject of angular momentum of light [8] and its transfer to matter [9,10] has added to the list of topics which can be studied via photo-induced electric responses in metals, in particular by considering currents induced by circularly polarized light [11,12]. The main goal of this paper is to provide general expressions for angular momentum transfer in plasmonic nanostructures and analyze its role in photo induced electric effects.

From classical point of view, photoinduced electric currents have two contributions resulting from (i) photogalvanic (or rectification) effect arising in non-centrosymmetric structures, and (ii) drag effect, which is due to absorption of electromagnetic quasimomentum by electrons (light pressure mechanism) [13,14]. The photogalvanic and drag effects are known for both linear and circular polarizations [14-17]. The circular photogalvanic effect produces currents according to $j_\alpha = i\kappa_{\alpha\beta}(\boldsymbol{E}\times\boldsymbol{E}^*)_\beta$ (where $\kappa_{\alpha\beta}$ are coefficients of proportionality) and is attributed to the absorption of circularly polarized light [18,19]. Generation of photoinduced currents in nonabsorbing media is classified as the inverse Faraday effect [20-24], and the produced currents are given by $\boldsymbol{j} \propto i\nabla\times(\boldsymbol{E}\times\boldsymbol{E}^*)$, corresponding to rectified magnetization $\boldsymbol{M} \propto i(\boldsymbol{E}\times\boldsymbol{E}^*)$. The phenomenological and microscopic models, and the comparison of the surface currents due to photogalvanic and photon drag effects in metal were considered in the situation of the anomalous skin effect by Gurevich *et al* [25,26].

Plasmon drag effect (PLDE) [11,12,27-37] is the giant enhancement of photoinduced electric currents in metal films and nanostructures in the conditions of surface plasmon resonance (SPR). This phenomenon attracts considerable attention due to its comprehensive nature [11,12,27-37], encompassing the topics of optical force distribution in matter [28,30,37], ultrafast optics [28], quantum plasmonics and hot electron kinetics [30]. It has been shown recently [30], that the PLDE induced current density can be found as $j = \frac{e}{m^*} t_{therm} f_{pr}$, where the timescale determining the PLDE is the thermalization time $t_{therm}$ of hot-electrons and $f_{pr}$ is the pressure force density acting on polarization charges in the metal [28]. Using the C-method theory [38], it was analytically shown that in a periodically modulated film, the work $W$ of the pressure force resulting in the electron drift in $x$ direction can be found as [30]

$$W = \overline{{f_{pr}}_x + \tan\theta \cdot {f_{pr}}_z d} = \sum_m \frac{\hbar k_m}{\hbar\omega}\overline{Q_m}\, d + \overline{(\tan\theta - a'){f_{pr}}_z} d - 2\pi\omega\, \mathrm{Im}\{\chi\}\overline{a''\Sigma_{SPP_y}} d \quad (1)$$

where bars denote averaging over a period, film thickness, and time, $\theta(x, y)$ is the position-dependent angle between electric current and the $x$-axis, $d$ is the period, $\chi$ is the susceptibility of the material, and $a(x)$ is the function describing the periodic height modulation. The first sum on the right-hand side of Eq. (1) is composed from contributions from different diffraction modes and is proportional to the energy absorption rate $\overline{Q_m}$ and the quasimomentum of each mode. The other two terms in Eq. (1) are not proportional to quasi-momentum and are related to the photogalvanic effect; the second term is due to non-laminar currents and includes the possible effects of surface considered by Gurevich et al. [25,26], whereas the third term is due to absorption of the SAM of SPPs $\boldsymbol{\Sigma}_{SPP} = \frac{1}{8\pi\omega}\mathrm{Im}\{\boldsymbol{E} \times \boldsymbol{E}^*\}$ and is related to the circular photogalvanic effect of Ivchenko, Pikus and Belinicher [18,19]. We call the contributions into plasmon-generated currents corresponding to the last two terms in Eq. (1) as *plasmogalvanic effects* (PLGE). Below we focus on PLGE related to the absorption of SAM of plasmons. We show that SAM of surface plasmon polaritons (SPPs) is absorbed by the metal plasma together with energy and linear momentum of SPPs (see Fig. 1(a)) and this SAM absorption leads to dramatic redistribution of SPP-induced forces on electrons and localizes them at the very surface of the metal in the thin atomic layer where the surface charges are concentrated (Fig. 1(b)). First, we present the general expressions for SAM transfer in matter and then consider effects in flat metal films in detail.

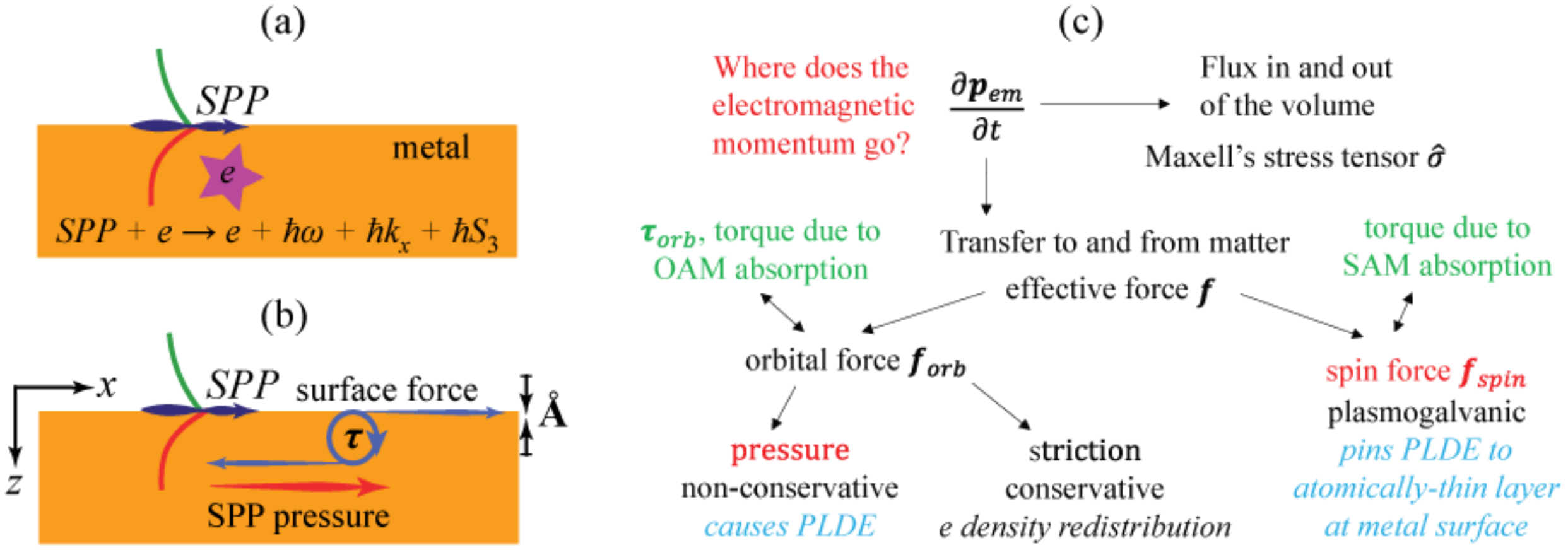

Fig. 1. (a) Schematic of the absorption of an SPP quantum by an electron involving transfer of energy $\hbar\omega$, momentum $\hbar k_x$ and average spin angular momentum $\hbar S_3$, where $S_3$ is the Stokes parameter, characterizing the helicity. (b) Schematic of the plasmon drag pinning to the atomic layer at the metal surface due to SAM absorption. The volume spin forces practically cancel SPP pressure and most of the SPP momentum is deposited by the surface force at the metal interface. (c) Channels for electromagnetic momentum transfer in media and corresponding photoinduced electric effects.

Let us start with the electromagnetic spin conservation law in a lossy material and show that the torque applied to a material by electromagnetic fields can be viewed as a dissipation term in the continuity equation for electromagnetic spin. The conservation law in a lossless dielectric was derived in Ref. [4]. Here, we adopt the following dual-symmetric definition of the spin angular momentum (SAM) of an electromagnetic field $\boldsymbol{\Sigma} = \boldsymbol{\Sigma}_e + \boldsymbol{\Sigma}_m$, where $\boldsymbol{\Sigma}_e = \frac{1}{4\pi c}\boldsymbol{E} \times \boldsymbol{A}$ and $\boldsymbol{\Sigma}_m = \frac{1}{4\pi c}\boldsymbol{H} \times \boldsymbol{C}$ [2,39], where the vector potentials are defined as $\boldsymbol{B} = \nabla \times \boldsymbol{A}$ and $\boldsymbol{D} = \nabla \times \boldsymbol{C}$ using the solenoidal nature of $\boldsymbol{D} = \boldsymbol{E} + 4\pi\boldsymbol{P}$ and $\boldsymbol{B} = \boldsymbol{H} + 4\pi\boldsymbol{M}$. The potentials are related to the fields as $\boldsymbol{E} = -\frac{1}{c}\frac{\partial \boldsymbol{A}}{\partial t}$ and $\boldsymbol{H} = \frac{1}{c}\frac{\partial \boldsymbol{C}}{\partial t}$. As we show in the Supplementary Material 1, the rate of the angular momentum transfer between the electromagnetic spin degree of freedom and matter satisfies the following continuity equation.

$$\frac{\partial \boldsymbol{\Sigma}}{\partial t} + \left(\nabla \cdot \hat{\delta}\right) = -\boldsymbol{\tau}, \qquad (2)$$

where $\hat{\delta}$ is the tensor of electromagnetic spin flux, analogous to Maxwell stress tensor for the flux of linear momentum of the electromagnetic field (cf. Eq. (3.24) of Ref. [4])

$$\hat{\delta} = \left\{\frac{1}{4\pi}\left(\boldsymbol{C} \otimes \boldsymbol{E} - \frac{(\boldsymbol{C}^c \cdot \boldsymbol{E})}{2}\hat{\mathrm{I}}\right) - \frac{1}{4\pi}\left(\boldsymbol{A} \otimes \boldsymbol{H} - \frac{(\boldsymbol{A}^c \cdot \boldsymbol{H})}{2}\hat{\mathrm{I}}\right)\right\}. \qquad (3)$$

and $\boldsymbol{\tau}$ is the torque volume density, which is composed from the torques associated with the interaction between the polarization and the electric fields $\boldsymbol{\tau_e}$ and the magnetization and the magnetic fields $\boldsymbol{\tau_m}$: $\boldsymbol{\tau} = \boldsymbol{\tau_e} + \boldsymbol{\tau_m}$.

$$\boldsymbol{\tau_e} = \frac{1}{c}\frac{\partial}{\partial t}(\boldsymbol{P} \times \boldsymbol{A}) + \boldsymbol{P} \times \boldsymbol{E}, \quad \boldsymbol{\tau_m} = \frac{1}{c}\frac{\partial}{\partial t}(\boldsymbol{M} \times \boldsymbol{C}) + \boldsymbol{M} \times \boldsymbol{H} \qquad (4)$$

In monochromatic fields the first terms in Eqs. (4) result in zero time-average torque, the time-averaged torque density applied to matter is

$$\bar{\boldsymbol{\tau}} = \frac{1}{2}\mathrm{Re}\{\boldsymbol{P} \times \boldsymbol{E}^* + \boldsymbol{M} \times \boldsymbol{H}^*\} \qquad (5)$$

However, in narrow-bandwidth pulses the first terms of Eqs. (4) lead to torque according to $\boldsymbol{\tau} = -2\pi\frac{d(\omega\chi_e)}{d\omega}\frac{\partial\overline{\boldsymbol{\Sigma}_e}}{\partial t} + 2\pi\frac{d(\omega\chi_m)}{d\omega}\frac{\partial\overline{\boldsymbol{\Sigma}_m}}{\partial t}$, (see Supplementary Material 2).

Both expressions for electric and magnetic torques (Eq. 5) can be understood as torques acting on dipole moments in electric and magnetic fields and are used, for example, in Landau-Lifshitz-Gilbert equation to describe magnetization dynamics [40,41]. The expression for the torque $\boldsymbol{\tau_e}$ applied to polarized matter has been used in the proposal of optical torque wrench devices [10], but the role of SAM absorption in matter has never been previously discussed (see also Refs. [42-46]). To show the relation of torque and SAM absorption explicitly, let one assume the material relationship of the form $\boldsymbol{P} = \chi_e\boldsymbol{E} + \xi_e\boldsymbol{H}$ and $= \chi_m\boldsymbol{H} + \xi_m\boldsymbol{E}$ , and the torque in Eq. (5) turns into

$$\bar{\boldsymbol{\tau}} = \frac{1}{2}\chi_e''\mathrm{Im}\{\boldsymbol{E} \times \boldsymbol{E}^*\} + \frac{1}{2}\chi_m''\mathrm{Im}\{\boldsymbol{H} \times \boldsymbol{H}^*\} + (\xi_m' - \xi_e')\,\mathrm{Re}\{\boldsymbol{E} \times \boldsymbol{H}^*\} + (\xi_e'' + \xi_m'')\,\mathrm{Im}\{\boldsymbol{E} \times \boldsymbol{H}^*\} =$$

$$= 4\pi\omega(\chi_e''\overline{\boldsymbol{\Sigma}_e} + \chi_m''\overline{\boldsymbol{\Sigma}_m}) + 8\pi(\xi_m' - \xi_e')\overline{\boldsymbol{S}} + 8\pi(\xi_e'' + \xi_m'')\tilde{\boldsymbol{S}} \qquad (6)$$

where $\overline{\boldsymbol{S}} = 1/(8\pi)\mathrm{Re}\{\boldsymbol{E} \times \boldsymbol{H}^*\}$ is the Poynting vector averaged over the optical period and $\tilde{\boldsymbol{S}} = 1/(8\pi)\mathrm{Im}\{\boldsymbol{E} \times \boldsymbol{H}^*\}$. Additionally, it appears that loss in bi-anisotropic media, or $\xi_m \neq \xi_e^*$ [47], can result in torque. This fact is of interest due to recent investigations of optical forces applied to bi-anisotropic particles [48-51].

In order to better understand the result of the SAM transfer from the electromagnetic field to matter we turn to linear momentum transfer (see Supplementary Material 3), which is described with the equation

$$\frac{\partial \boldsymbol{p}}{\partial t} + \nabla \cdot \hat{\sigma} = -\boldsymbol{f},$$

where $\boldsymbol{p} = 1/(4\pi c)(\boldsymbol{E} \times \boldsymbol{H})$ is the linear electromagnetic momentum , $\hat{\sigma}$ is the Maxwell stress tensor [52], and $\boldsymbol{f}$ is the volume density of the effective force acting in the medium.

The force has an interesting structure, reminiscent of both Lorentz and Einstein-Laub forces [44-46,53-58], and is identical to the Lorentz force for $\boldsymbol{M} = 0$, which was first applied to PLDE (considering only electric responses) in Refs. [28,30].

Considering separately volume and surface contributions, the volume density of the force $\boldsymbol{f}$ is

$$\boldsymbol{f} = -(\nabla\cdot\boldsymbol{P})\boldsymbol{E} + \frac{1}{c}\frac{\partial \boldsymbol{P}}{\partial t}\times\boldsymbol{H} - (\nabla\cdot\boldsymbol{M})\boldsymbol{H} - \frac{1}{c}\frac{\partial \boldsymbol{M}}{\partial t}\times\boldsymbol{E} = \boldsymbol{f}_A + \boldsymbol{f}_{orb}^{e} + \boldsymbol{f}_{orb}^{m} + \boldsymbol{f}_{spin}^{ve} + \boldsymbol{f}_{spin}^{vm} \quad (7)$$

where

(Abraham force) $\boldsymbol{f}_A = \frac{1}{c}\frac{\partial}{\partial t}(\boldsymbol{P}\times\boldsymbol{H} - \boldsymbol{M}\times\boldsymbol{E} + 4\pi\boldsymbol{P}\times\boldsymbol{M})$ (7a)

(orbital electric force) $\boldsymbol{f}_{orb}^{e} = \nabla_{\boldsymbol{E}}(\boldsymbol{P}^c\cdot\boldsymbol{E})$ (7b)

(orbital magnetic force) $\boldsymbol{f}_{orb}^{m} = \nabla_{\boldsymbol{H}}(\boldsymbol{M}^c\cdot\boldsymbol{H})$ (7c)

(spin electric volume force) $\boldsymbol{f}_{spin}^{ve} = -\nabla\cdot(\boldsymbol{P}\otimes\boldsymbol{E})$ (7d)

(spin magnetic volume force) $\boldsymbol{f}_{spin}^{vm} = -\nabla\cdot(\boldsymbol{M}\otimes\boldsymbol{H})$ (7e)

The surface density (after integration of the volume density Eq. (7) across the metal interface) is given by

$$\boldsymbol{f}_s = (\boldsymbol{P}\cdot\hat{\boldsymbol{n}})\boldsymbol{E} + (\boldsymbol{M}\cdot\hat{\boldsymbol{n}})\boldsymbol{H} = \boldsymbol{f}_{spin}^{se} + \boldsymbol{f}_{spin}^{sm} \quad (8)$$

(spin electric volume force) $\boldsymbol{f}_{spin}^{se} = (\boldsymbol{P}\cdot\hat{\boldsymbol{n}})\boldsymbol{E}$ (8a)

(spin magnetic volume force) $\boldsymbol{f}_{spin}^{sm} = (\boldsymbol{M}\cdot\hat{\boldsymbol{n}})\boldsymbol{H}$ (8b)

Please notice the symmetry of the force Eq. (7)-(8) with respect to electric and magnetic responses of materials. The Abraham force [Eq. (7a)] is zero in monochromatic fields, similarly to the first terms in Eqs. (4) or to the electromagnetic energy in media [59], and for pulses in $\boldsymbol{M} = 0$ media was considered in [60]. The forces in Eqs. (7b) and (7c) can be viewed as the force responsible for linear momentum transfer and change of the orbital angular momentum [1-4] (i.e. OAM transfer with torque $\boldsymbol{\tau}_{orb} = \boldsymbol{r}\times\boldsymbol{f}_{orb}$ ) and can be referred to as the orbital force $\boldsymbol{f}_{orb} = \nabla_{\boldsymbol{E}}(\boldsymbol{P}^c\cdot\boldsymbol{E}) + \nabla_{\boldsymbol{H}}(\boldsymbol{M}^c\cdot\boldsymbol{H})$. The rectified part of the orbital force can be represented as the sum of striction and pressure forces [28]

$$\overline{\boldsymbol{f}_{orb}^{e}} = \frac{1}{2}\mathrm{Re}\{\nabla_{\boldsymbol{E}}(\boldsymbol{P}^c\cdot\boldsymbol{E}^*)\} = \frac{1}{4}\chi_e'\nabla(|\boldsymbol{E}|^2) - \frac{1}{2}\chi_e''\mathrm{Im}\{\nabla_{\boldsymbol{E}}(\boldsymbol{E}^c\cdot\boldsymbol{E}^*)\}, \quad (9a)$$

$$\overline{\boldsymbol{f}_{orb}^{m}} = \frac{1}{2}\mathrm{Re}\{\nabla_{\boldsymbol{H}}(\boldsymbol{M}^c\cdot\boldsymbol{H}^*)\} = \frac{1}{4}\chi_m'\nabla(|\boldsymbol{H}|^2) - \frac{1}{2}\chi_m''\mathrm{Im}\{\nabla_{\boldsymbol{H}}(\boldsymbol{H}^c\cdot\boldsymbol{H}^*)\}, \quad (9b)$$

The pressure force (the terms with $\chi''$ in Eqs. (9)) is proportional to the wave vector of the electromagnetic field and is the source of the PLDE. In Ref. [30] it was demonstrated that the linear momentum transfer is directly related to energy absorption, which is manifested in PLDE experiments [27,29]. The remaining contributions in Eqs. (7)-(8) can be classified as the spin force, whose volume density is $\boldsymbol{f}_{spin}^{v} = \boldsymbol{f}_{spin}^{ve} + \boldsymbol{f}_{spin}^{vm}$ and surface density is $\boldsymbol{f}_{spin}^{s} = \boldsymbol{f}_s = \boldsymbol{f}_{spin}^{se} + \boldsymbol{f}_{spin}^{sm}$. The electric part of $\boldsymbol{f}_{spin}$ was first obtained in Ref. [28] (for $\boldsymbol{M} = 0$).

The spin force does not directly contribute to the PLDE effect; nevertheless, it significantly affects the distribution of the net force in the material and leads to plasmogalvanic effects as we discuss below. Indeed, as was shown in [28] the force densities in (7d) and (8a) satisfy $\oint \boldsymbol{f}_{spin}^{se}\cdot ds - \int \boldsymbol{f}_{spin}^{ve}\cdot dV = 0$. Similarly, $\oint \boldsymbol{f}_{spin}^{sm}\cdot ds - \int \boldsymbol{f}_{spin}^{vm}\cdot dV = 0$. Despite the zero overall contribution into the total force, these forces result in the torque of Eqs. (5)-(6). The torques of Eq. (5) can be presented as (see Supplementary Materials 4)

$$\int(\boldsymbol{P}\times\boldsymbol{E})dV=\oint(\boldsymbol{r}\times\boldsymbol{f}_{spin}^{se})ds+\int(\boldsymbol{r}\times\boldsymbol{f}_{spin}^{ve})dV, \tag{10a}$$

$$\int(\boldsymbol{M}\times\boldsymbol{H})dV=\oint(\boldsymbol{r}\times\boldsymbol{f}_{spin}^{sm})ds+\int(\boldsymbol{r}\times\boldsymbol{f}_{spin}^{vm})dV. \tag{10b}$$

Note that despite the fact that the SAM absorption torque is only proportional to $\chi''$ (see Eq. (6)), the corresponding rectified spin forces $\boldsymbol{f}_{spin}$ have both $\chi'$ and $\chi''$ contributions

$$\overline{\boldsymbol{f}_{spin}^{ve}}=-\frac{1}{2}\mathrm{Re}\{\nabla\cdot(\boldsymbol{P}\otimes\boldsymbol{E}^*)\}=\frac{1}{4}\chi_e''\mathrm{Im}\{\nabla\times(\boldsymbol{E}\times\boldsymbol{E}^*)\}-\frac{1}{2}\chi_e'\mathrm{Re}\{\nabla\cdot(\boldsymbol{E}\otimes\boldsymbol{E}^*)\} \tag{11a}$$

$$\overline{\boldsymbol{f}_{spin}^{vm}}=-\frac{1}{2}\mathrm{Re}\{\nabla\cdot(\boldsymbol{M}\otimes\boldsymbol{H}^*)\}=\frac{1}{4}\chi_m''\mathrm{Im}\{\nabla\times(\boldsymbol{H}\times\boldsymbol{H}^*)\}-\frac{1}{2}\chi_m'\mathrm{Re}\{\nabla\cdot(\boldsymbol{H}\otimes\boldsymbol{H}^*)\} \tag{11b}$$

The first terms of Eq. (11) $\frac{1}{4}\chi_e''\mathrm{Im}\{\nabla\times(\boldsymbol{E}\times\boldsymbol{E}^*)\}$ and $\frac{1}{4}\chi_m''\mathrm{Im}\{\nabla\times(\boldsymbol{H}\times\boldsymbol{H}^*)\}$ can be understood as absorption of the spin part of the Poynting current $P_{3sp}=\mathrm{Im}\{\nabla\times(\boldsymbol{E}\times\boldsymbol{E}^*)\}$ from Ref. [1].

Now let us consider the results of SAM transfer for a simple case of SPP propagation at a flat metal-dielectric interface with $\mu=1$. The fields of a SPP in the metal ($z>0$) are

$$\boldsymbol{H}=\hat{\boldsymbol{y}}\,e^{-\xi z}e^{i(kx-\omega t)},\quad \boldsymbol{E}=\frac{1}{k_0\varepsilon}(-\hat{\boldsymbol{z}}k_x+\hat{\boldsymbol{x}}i\xi)e^{-\xi z}e^{i(kx-\omega t)},\frac{\xi}{\varepsilon_m}=-\frac{\kappa_d}{\varepsilon_d}. \tag{12}$$

The last equation is the condition of SPP existence, ensuring the matching of longitudinal electric fields at the metal-dielectric interface, which leads to the dispersion $k_{SPP}(\omega)$ of SPPs (see Fig. 2(a)).

Inside the metal the volume density of the force acting on electron is a combination of the orbital and spin forces $\overline{\boldsymbol{f}}=\overline{\boldsymbol{f}_{orb}}+\overline{\boldsymbol{f}_{spin}}$. In the orbital force only the pressure force is doing work on electrons, while striction is a potential force and is responsible for redistribution of electron density, but not photoinduced currents [30]. The spin part of the force inside the metal is

$$\overline{\boldsymbol{f}_{spin}}=-\frac{1}{2}\mathrm{Re}\{\nabla\cdot(\boldsymbol{P}\otimes\boldsymbol{E}^*)\}=-\frac{1}{2}\mathrm{Re}\{(\boldsymbol{P}\cdot\nabla)\boldsymbol{E}^*\}=-\frac{1}{2}\mathrm{Re}\{\partial_z(P_z\boldsymbol{E}^*)\}-\frac{1}{2}\mathrm{Re}\{\partial_x(P_x\boldsymbol{E}^*)\} \tag{13}$$

Compare this volume force density in the metal with the force surface density on the surface of the metal $\overline{\boldsymbol{f}_{spin}^{se}}=\frac{1}{2}\mathrm{Re}\{P_z\boldsymbol{E}^*\}\Big|_{z=0}$ (see Eq. (8a)). The first term in the volume spin force integrated over the cross-section of the metal perpendicular to plasmon propagation gives exactly the opposite of the surface force. The second term integrated over the cross-section of the metal in the direction of SPP propagation is equal to zero assuming fields decay out for $x\to\pm\infty$. If the fields do not fully decay, this part of the spin force is opposite to the surface force created at the ends of the metal in the *x*-direction. In any case the second term is due to the decay of SPPs in the direction of propagation and does no overall work on electrons in the case of the laminar current [30]. For this reason, below we only consider one part of the spin force $\overline{\boldsymbol{f}_{spin_z}}=-\frac{1}{2}\mathrm{Re}\{\partial_z(P_z\boldsymbol{E}^*)\}$ and disregard $\overline{\boldsymbol{f}_{spin_x}}=-\frac{1}{2}\mathrm{Re}\{\partial_x(P_x\boldsymbol{E}^*)\}$.

The SAM absorption is related to the action of spin forces in volume and surface according to Eqs. (10) and the corresponding torque of these forces in the SPP field [Eq. (12)] can be found using Eq. (6) as $\overline{\boldsymbol{\tau}}=-\chi''\mathrm{Im}\{E_xE_z^*\}\hat{\boldsymbol{y}}$. It is important to compare this torque with the energy absorption rate $\bar{Q}=-\frac{\omega}{2}\mathrm{Im}\{P_\alpha E_\alpha^*\}$ as was done in Ref. [30] for linear momentum. We show in the Supplementary Material 5 that the relationship between the torque and energy absorption rate is $\overline{\boldsymbol{\tau}}=\frac{\hbar S_3}{\hbar\omega}\bar{Q}$, where $S_3=\frac{2\xi k_x}{(k_x^2+\xi^2)}$ is the Stokes parameter, characterizing the helicity (or the degree of circular polarization) in electromagnetic field [6,7,61]. Note that here we are comparing energy transferred to matter with the corresponding torque, and not the energy and SAM content in the field as in Refs [6,7,61]. Considering the results of Ref. [30] this means that with absorption of SPP energy quantum $\hbar\omega$ and momentum quantum $\hbar k_x$ electrons on average gain $\hbar S_3$ amount of angular momentum due to absorption of SAM (see Fig. 1(a)).

In principle, the total force acting on electrons can be calculated using just the pressure part of the orbital force [30,37], however, the inclusion of the spin force into consideration will provide information of the force distribution and role of the surface charge layer. The study of the surface charge layer between metal and dielectric has been a major direction in research of metal non-locality [62,63] and has attracted a lot of attention recently [64]. It has been noted that non-local effects at the metal-dielectric interface be approximated by introduction of an anisotropic transition layer [65]. Here we do not pursue the goal of modeling the non-locality as such, but would like to propose a toy-model of a transitional metal surface layer, which gives an idea of how the SAM-absorption torque affects the momentum transfer from SPPs to metal plasma.

We first note that at a metal-dielectric interface the tangential component of electric field $E_x$ is continuous, while $D_x$ is not continuous and changes sign through the interface due to negativity of the dielectric permittivity of metal $\varepsilon_m$. This implies that the dielectric permittivity passes through an epsilon-near-zero (ENZ) transition at the metal-dielectric interface in the longitudinal direction. At the same time the normal component of the electric field $E_z$ has a discontinuity at the interface, such that $D_z$ is continuous. This means that the transverse to the interface dielectric permittivity exhibits an epsilon-near-pole (ENP) transition at the metal boundary. The study of ENZ-ENP metasurfaces has been started in a recent publication [66], where they were shown to represent the ultimate limit for wave plate thickness. Here we use the concept of *gradient ENZ-ENP metasurface* to model a transition layer at the metal-dielectric boundary. We assume that at this boundary the metal fraction $f(z)$ is gradually changing from 0 to 1, such that a distributed ENZ-ENP metasurface has dielectric permittivities, which depend on normal coordinate $z$ as $\varepsilon_x(z) = \varepsilon_m f(z) + \varepsilon_d\big(1 - f(z)\big)$ and $\varepsilon_z^{-1}(z) = \varepsilon_m^{-1} f(z) + \varepsilon_d^{-1}\big(1 - f(z)\big)$.

We use the following function for $f(z) = \frac{1}{2} + \frac{1}{2\alpha d}(\ln(\cosh \alpha z) - \ln(\cosh \alpha(d - z)))$, which is shown in Fig. 2(c) for our calculations (we use $d = 2$ Å and $\alpha = 10$ Å$^{-1}$). The TM polarized fields of SPP wave at the metal-dielectric boundary with such a metasurface can be written as $H_y = H_y(z)e^{i(kx-\omega t)}$, where function $H_y(z)$ satisfies the following equation in the transition layer

$$-\frac{1}{\varepsilon_x} H_y'' - H_y' \cdot \partial_z\left(\frac{1}{\varepsilon_x}\right) = \left(k_0^2 - \frac{k^2}{\varepsilon_z}\right) H_y \quad . \tag{14}$$

We solve this equation and match the results at the boundaries of the ENZ-ENP metasurface to get the wave vector of the resulting SPPs (see the dots in Fig. 2(a), which follow very closely the dispersion of SPPs in local model Eq. (13)). After that we obtain the magnetic field distribution given by yellow line in Fig. 2(b) and compare it with the local model (cyan dashed line), and the electric field distribution (see Fig. 2(b)-(c)) according to $E_z = -\frac{kH_y}{k_0\varepsilon_z}$ and $E_x = -\frac{i}{k_0}\frac{H_y'}{\varepsilon_x}$. The normal to the surface component $E_z$ is shown on the nanometric scale in Fig. 2(b) in blue and match the local model (magenta dashed curve).

In Fig. 2(c) one can see the gradual transition of $E_z$ from the metal value to the value in the air within the 2 Å metasurface (blue curve) and compare it to the abrupt jump in the local model. In the metasurface model the oscillating surface charge $\sigma(z) \approx -P_z(z=0)/d$ in the SPP excitation is distributed over the metasurface as seen in Fig. 2(c). Its value corresponds to the local model with $\sigma = -P_z$.

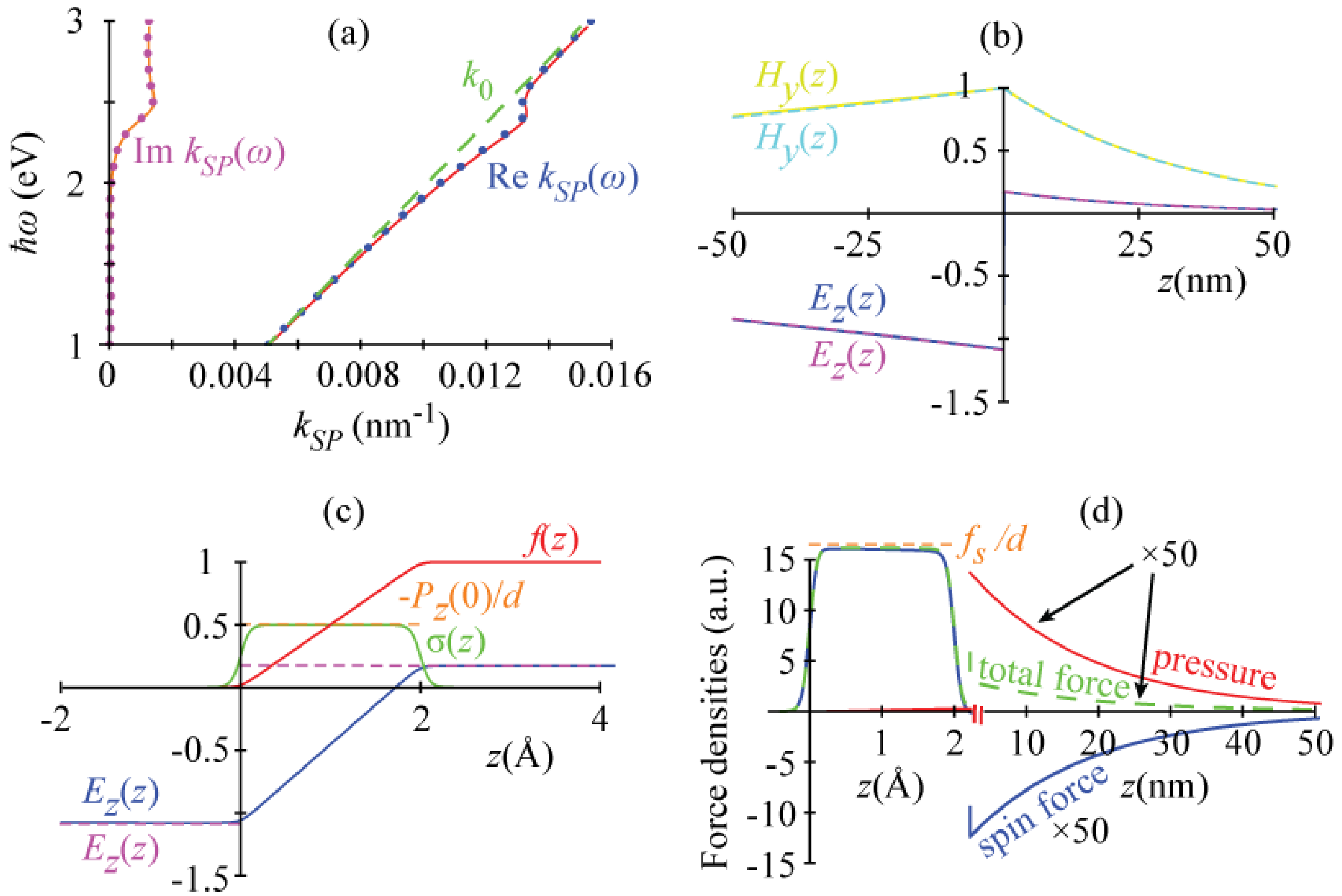

Fig. 2. (a) SPP dispersion in the local model (continuous curves, Eq. (13)) and the ENZ-ENP metasurface model (dots). (b) Comparison of the field distributions in the local model (shown in yellow and blue color) and metasurface approximation (dashed cyan and purple) on the nanoscopic scale. (c) The fields and surface charges in the atomic metasurface region. The metal fraction function $f(z)$ is shown in red. The normal to the interface component of the electric field $E_z$ is shown in blue (metasurface model) and dashed magenta (local model). The surface charge density $\sigma(z)$ is shown in green (metasurface model) and dashed orange (local model). (d) The effect of the spin force (blue) on the total momentum transfer from SPP to electrons (dashed green) as compared to considering only the plasmon pressure force (red). For comparison the surface force in the local model is shown in orange.

Introduction of the ENZ-ENP metasurface model allows us to visualize the pinning of the plasmon drag to the atomic layer at the metal surface. In Fig. 2(d) we first plot the spin force $\overline{\boldsymbol{f}_{spin_z}}$ (blue) within the metasurface $0 < z < 2$ Å and compare it to the surface force $\overline{\boldsymbol{f}_s}$ (Eq. (8a)) divided by 2 Å (orange dashed line). One can see the obvious similarity of the results of the metasurface model and the local model, and at the same time more physical insight can be drawn from the calculations with the metasurface. The good correspondence between the metasurface model and the local model in calculation of the surface force means that the local model can be successfully applied to include plasmonic SAM absorption in the future calculations. On the nanometric scale (shown to the right of the $z$-axis break after the metasurface region) the spin force is also shown in blue and multiplied by 50. Integration of the blue curve in Fig. 2(d) over $z$ results in no overall force in accordance with Eq. (10a). Nevertheless, inclusion of the spin force considerably changes the force distribution towards the interface of the metal and pins it to the atomic metasurface region. The total PLDE can be calculated by integrating only the pressure force over the volume of the metal [30,37]. The plasmonic pressure distribution is shown in red in Fig. 2(d) and exponentially decays within the skin layer of the metal (multiplied by 50 on the nanometric scale). The total force on electrons including the spin force is shown as green dashed line and closely follows the surface component of the spin force in the atomic metasurface region, but is 4 times weaker than pressure in the metal volume (multiplied by 50 on the nanometric scale), which demonstrates the drag pinning.

In the recent years the plasmonics field has experienced a considerable shift towards atomic scale features, structures [67-69], gaps or cavities [70,71], surface charge nonlinearities and photoinduced voltages [72,31] or atomic-layer 2d materials [73-75]. For example, plasmons propagating along graphene structures exhibit dispersion, electromagnetic confinement and loss properties that can be compared to those of plasmons at noble metal interfaces [76]. Main difference between plasmon excitations in these materials is that the optical response of graphene is more sensitive to electrostatic interactions in comparison to metals, due to obvious differences in electronic confinement, which can find a plethora of exciting applications [77,13]. Nevertheless, as we show in this manuscript, the sensitivity of noble-metal plasmons to the environment at the metal-dielectric interface could be considerably stronger than was previously deemed. The induced charge in the noble-metal plasmons, as in the case of graphene plasmons, is localized in an atomically-thin layer at the very edge of the metal, despite of the fact that electrons can travel into the bulk of the metal from the interface [62-64]. We demonstrate that, due to the absorption of the spin angular momentum of plasmons, most of the momentum dissipation from plasmons to one-particle electronic excitations occurs at the location of this atomically-thin induced charge layer (see Fig. 1(b) for schematic). This should lead to rectified surface currents, which we expect to be sensitive to the environment at the metal-dielectric interface, but which are at the same time more topologically robust than in the atomic-layer materials, due to the possibility to shift electron states into the bulk of the metal.

To conclude, we have considered the SAM transfer between electromagnetic fields and matter. The corresponding spin forces in the fields of SPPs at metal-dielectric interfaces lead to pinning of the plasmon drag linear momentum transfer to the atomically-thin layers at the metal interface.

**Supplementary Material**

**Part 1.** Consider the spin angular momentum (SAM) of plasmonic field

$$\boldsymbol{\Sigma} = \frac{1}{4\pi c}(\boldsymbol{E}\times\boldsymbol{A} + \boldsymbol{H}\times\boldsymbol{C}), \text{ where } = \nabla\times\boldsymbol{C},\ \boldsymbol{B} = \nabla\times\boldsymbol{A}\,.$$

$$\nabla\times\boldsymbol{E} = -\frac{1}{c}\frac{\partial\boldsymbol{B}}{\partial t} = -\frac{1}{c}\frac{\partial(\nabla\times\boldsymbol{A})}{\partial t}, \qquad \boldsymbol{E} = -\frac{1}{c}\frac{\partial\boldsymbol{A}}{\partial t}$$

$$\nabla\times\boldsymbol{H} = \frac{1}{c}\frac{\partial\boldsymbol{D}}{\partial t} = \frac{1}{c}\frac{\partial(\nabla\times\boldsymbol{C})}{\partial t}, \qquad \boldsymbol{H} = \frac{1}{c}\frac{\partial\boldsymbol{C}}{\partial t}$$

Below we derive the continuity equation (2)

$$\frac{\partial\boldsymbol{\Sigma}}{\partial t} = \frac{1}{4\pi c}\frac{\partial}{\partial t}(\boldsymbol{E}\times\boldsymbol{A} + \boldsymbol{H}\times\boldsymbol{C}) = \frac{1}{4\pi c}\left(\frac{\partial\boldsymbol{E}}{\partial t}\times\boldsymbol{A} + \frac{\partial\boldsymbol{H}}{\partial t}\times\boldsymbol{C}\right) + \frac{1}{4\pi c}\left(\boldsymbol{E}\times\frac{\partial\boldsymbol{A}}{\partial t} + \boldsymbol{H}\times\frac{\partial\boldsymbol{C}}{\partial t}\right)$$

$$= \frac{1}{4\pi c}\left(\left(c\nabla\times\boldsymbol{H} - 4\pi\frac{\partial\boldsymbol{P}}{\partial t}\right)\times\boldsymbol{A} + \left(-c\nabla\times\boldsymbol{E} - 4\pi\frac{\partial\boldsymbol{M}}{\partial t}\right)\times\boldsymbol{C}\right)$$

$$= \frac{1}{4\pi}(-(\nabla\times\boldsymbol{E})\times\boldsymbol{C} + (\nabla\times\boldsymbol{H})\times\boldsymbol{A}) - \frac{1}{c}\left(\frac{\partial\boldsymbol{P}}{\partial t}\times\boldsymbol{A} + \frac{\partial\boldsymbol{M}}{\partial t}\times\boldsymbol{C}\right) = -(\nabla\cdot\hat{\delta}) - \boldsymbol{\tau}$$

$$\hat{\delta}_{ik} = -\frac{1}{4\pi}(-C_i E_k + A_i H_k) + \frac{1}{8\pi}\big(-(\boldsymbol{C}^c \cdot \boldsymbol{E}) + (\boldsymbol{A}^c \cdot \boldsymbol{H})\big)\delta_{ik}$$
$$= \left\{\frac{1}{4\pi}\left(\boldsymbol{C} \otimes \boldsymbol{E} - \frac{(\boldsymbol{C}^c \cdot E)}{2}\hat{I}\right) - \frac{1}{4\pi}\left(\boldsymbol{A} \otimes \boldsymbol{H} - \frac{(\boldsymbol{A}^c \cdot \boldsymbol{H})}{2}\hat{I}\right)\right\}_{ik}$$

Torque density acting on the medium can be written as

$$\boldsymbol{\tau} = \frac{1}{c}\left(\frac{\partial \boldsymbol{P}}{\partial t} \times \boldsymbol{A} + \frac{\partial \boldsymbol{M}}{\partial t} \times \boldsymbol{C}\right) = \frac{1}{c}\frac{\partial}{\partial t}(\boldsymbol{P} \times \boldsymbol{A} + \boldsymbol{M} \times \boldsymbol{C}) - \frac{1}{c}\left(\boldsymbol{P} \times \frac{\partial \boldsymbol{A}}{\partial t} + \boldsymbol{M} \times \frac{\partial \boldsymbol{C}}{\partial t}\right)$$
$$= \frac{1}{c}\frac{\partial}{\partial t}(\boldsymbol{P} \times \boldsymbol{A} + \boldsymbol{M} \times \boldsymbol{C}) + \boldsymbol{P} \times \boldsymbol{E} + \boldsymbol{M} \times \boldsymbol{H}$$

**Part 2.** Now let us discuss the electromagnetic torque in dispersive transparent media, which can be considered in a similar manner to the Brillouin internal energy stored in electromagnetic field in media [59] or Abraham force [59,60]. The second terms in Eqs. (4) have the same form as in the monochromatic fields, and we focus on the first terms, which are zero in monochromatic fields but play a role in a narrow bandwidth pulse [59]. In this case $\frac{\partial \boldsymbol{P}}{\partial t} = -i\omega\chi_e(\omega)\boldsymbol{E} + \frac{d(\omega\chi_e)}{d\omega}\frac{\partial \boldsymbol{E}_0}{\partial t}e^{-i\omega t}$ and $\frac{\partial \boldsymbol{M}}{\partial t} = -i\omega\chi_m(\omega)\boldsymbol{H} + \frac{d(\omega\chi_m)}{d\omega}\frac{\partial \boldsymbol{H}_0}{\partial t}e^{-i\omega t}$, $\boldsymbol{E}_0(t)$ and $\boldsymbol{H}_0(t)$ are slowly varying amplitudes of the fields. In this situation $\boldsymbol{\tau} = \frac{1}{c}\frac{\partial}{\partial t}(\boldsymbol{P} \times \boldsymbol{A} + \boldsymbol{M} \times \boldsymbol{C}) =$

$$= \frac{1}{4c}\left(\frac{\partial \boldsymbol{P}^*}{\partial t} \times \boldsymbol{A} + \frac{\partial \boldsymbol{P}}{\partial t} \times \boldsymbol{A}^* + \frac{\partial \boldsymbol{M}^*}{\partial t} \times \boldsymbol{C} + \frac{\partial \boldsymbol{M}}{\partial t} \times \boldsymbol{C}^*\right) - \frac{1}{2}\mathrm{Re}\{\boldsymbol{P} \times \boldsymbol{E}^* - \boldsymbol{M} \times \boldsymbol{H}^*\} =$$

$$= \frac{1}{4c}\frac{d(\omega\chi_e)}{d\omega}\left(\frac{\partial \boldsymbol{E}_0^*}{\partial t} \times \left(-i\frac{c}{\omega}\boldsymbol{E}_0\right) + \frac{\partial \boldsymbol{E}_0}{\partial t} \times \left(i\frac{c}{\omega}\boldsymbol{E}_0^*\right)\right) + \frac{1}{4c}\frac{d(\omega\chi_m)}{d\omega}\left(\frac{\partial \boldsymbol{H}_0^*}{\partial t} \times \left(i\frac{c}{\omega}\boldsymbol{H}_0\right) + \frac{\partial \boldsymbol{H}_0}{\partial t} \times \left(-i\frac{c}{\omega}\boldsymbol{H}_0^*\right)\right) =$$

$$= -\frac{1}{4\omega}\frac{d(\omega\chi_e)}{d\omega}\frac{\partial}{\partial t}\mathrm{Im}\{\boldsymbol{E} \times \boldsymbol{E}^*\} + \frac{1}{4\omega}\frac{d(\omega\chi_m)}{d\omega}\frac{\partial}{\partial t}\mathrm{Im}\{\boldsymbol{H} \times \boldsymbol{H}^*\} = -2\pi\frac{d(\omega\chi_e)}{d\omega}\frac{\partial \overline{\boldsymbol{\Sigma}_e}}{\partial t} + 2\pi\frac{d(\omega\chi_m)}{d\omega}\frac{\partial \overline{\boldsymbol{\Sigma}_m}}{\partial t}$$

**Part 3.** The continuity equation for the electromagnetic linear momentum $\boldsymbol{p} = 1/(4\pi c)(\boldsymbol{E} \times \boldsymbol{H})$ can be found as

$$\frac{\partial \boldsymbol{p}}{\partial t} = \frac{1}{4\pi c}\left(\frac{\partial \boldsymbol{E}}{\partial t} \times \boldsymbol{H} + \boldsymbol{E} \times \frac{\partial \boldsymbol{H}}{\partial t}\right) =$$

$$= \frac{1}{4\pi c}\left(\frac{\partial \boldsymbol{D}}{\partial t} \times \boldsymbol{H} + \boldsymbol{E} \times \frac{\partial \boldsymbol{B}}{\partial t}\right) - \frac{1}{c}\left(\frac{\partial \boldsymbol{P}}{\partial t} \times \boldsymbol{H} + \boldsymbol{E} \times \frac{\partial \boldsymbol{M}}{\partial t}\right) + \frac{1}{4\pi}(\nabla \cdot \boldsymbol{D})\boldsymbol{E} + \frac{1}{4\pi}(\nabla \cdot \boldsymbol{B})\boldsymbol{H}$$

$$= \frac{1}{4\pi}\big((\boldsymbol{\nabla} \times \boldsymbol{H}) \times \boldsymbol{H} + (\boldsymbol{\nabla} \times \boldsymbol{E}) \times \boldsymbol{E} + (\nabla \cdot \boldsymbol{E})\boldsymbol{E} + (\nabla \cdot \boldsymbol{H})\boldsymbol{H}\big)$$

$$+(\nabla \cdot \boldsymbol{P})\boldsymbol{E} - \frac{1}{c}\frac{\partial \boldsymbol{P}}{\partial t} \times \boldsymbol{H} + \frac{1}{4\pi}(\nabla \cdot \boldsymbol{M})\boldsymbol{H} + \frac{1}{c}\frac{\partial \boldsymbol{M}}{\partial t} \times \boldsymbol{E}$$

which using, $(\boldsymbol{\nabla} \times \boldsymbol{X}) \times \boldsymbol{Y} + (\nabla \cdot \boldsymbol{Y})\boldsymbol{X} = -\nabla_{\boldsymbol{X}}(\boldsymbol{X} \cdot \boldsymbol{Y}) + \nabla \cdot (\boldsymbol{Y} \otimes \boldsymbol{X})$, can be written as

$$\frac{\partial \boldsymbol{p}}{\partial t} + \nabla \cdot \hat{\sigma} = -\boldsymbol{f}$$

$$\nabla \cdot \hat{\sigma} = \frac{1}{4\pi}(-(\nabla \cdot \boldsymbol{E})\boldsymbol{E} - (\nabla \cdot \boldsymbol{H})\boldsymbol{H} - (\boldsymbol{\nabla} \times \boldsymbol{H}) \times \boldsymbol{H} - (\boldsymbol{\nabla} \times \boldsymbol{E}) \times \boldsymbol{E}) =$$

$$= \frac{1}{4\pi}\left(\nabla_{\boldsymbol{H}}(\boldsymbol{H}\cdot\boldsymbol{H}) - \nabla\cdot(\boldsymbol{H}\otimes\boldsymbol{H}) + \nabla_{\boldsymbol{E}}(\boldsymbol{E}\cdot\boldsymbol{E}) - \nabla\cdot(\boldsymbol{E}\otimes\boldsymbol{E})\right)$$

$$= \frac{1}{4\pi}\nabla\cdot\left(\frac{E^2+H^2}{2}\hat{\boldsymbol{I}} + (\boldsymbol{E}\otimes\boldsymbol{E}) + (\boldsymbol{H}\otimes\boldsymbol{H})\right)$$

where $\hat{\sigma}$ is the Maxwell stress tensor [52], and the volume density of force describing linear momentum transfer between an electromagnetic field and matter is

$$\boldsymbol{f} = -(\nabla\cdot\boldsymbol{P})\boldsymbol{E} + \frac{1}{c}\frac{\partial\boldsymbol{P}}{\partial t}\times\boldsymbol{H} - (\nabla\cdot\boldsymbol{M})\boldsymbol{H} - \frac{1}{c}\frac{\partial\boldsymbol{M}}{\partial t}\times\boldsymbol{E}.$$

Integrating over an interface between two materials we get the surface density of this force

$$\boldsymbol{f}_s = (\boldsymbol{P}\cdot\hat{\boldsymbol{n}})\boldsymbol{E} + (\boldsymbol{M}\cdot\hat{\boldsymbol{n}})\boldsymbol{H},$$

where $\hat{\boldsymbol{n}}$ is the normal to the metal boundary pointing outwards. The force $\boldsymbol{f}$ can be rewritten as

$$\boldsymbol{f} = -(\nabla\cdot\boldsymbol{P})\boldsymbol{E} + \frac{1}{c}\frac{\partial}{\partial t}(\boldsymbol{P}\times\boldsymbol{H}) - \frac{1}{c}\boldsymbol{P}\times\frac{\partial\boldsymbol{H}}{\partial t} - (\nabla\cdot\boldsymbol{M})\boldsymbol{H} - \frac{1}{c}\frac{\partial}{\partial t}(\boldsymbol{M}\times\boldsymbol{E}) + \frac{1}{c}\boldsymbol{M}\times\frac{\partial\boldsymbol{E}}{\partial t}$$

$$= -(\nabla\cdot\boldsymbol{P})\boldsymbol{E} + \frac{1}{c}\frac{\partial}{\partial t}(\boldsymbol{P}\times\boldsymbol{H}) - \frac{1}{c}\boldsymbol{P}\times\frac{\partial\boldsymbol{B}}{\partial t} + \frac{4\pi}{c}\boldsymbol{P}\times\frac{\partial\boldsymbol{M}}{\partial t} - (\nabla\cdot\boldsymbol{M})\boldsymbol{H} - \frac{1}{c}\frac{\partial}{\partial t}(\boldsymbol{M}\times\boldsymbol{E}) + \frac{1}{c}\boldsymbol{M}\times\frac{\partial\boldsymbol{D}}{\partial t} - \frac{4\pi}{c}\boldsymbol{M}\times\frac{\partial\boldsymbol{P}}{\partial t}$$

$$= -(\nabla\cdot\boldsymbol{P})\boldsymbol{E} + \boldsymbol{P}\times(\nabla\times\boldsymbol{E}) - (\nabla\cdot\boldsymbol{M})\boldsymbol{H} + \boldsymbol{M}\times(\nabla\times\boldsymbol{H}) + \frac{1}{c}\frac{\partial}{\partial t}(\boldsymbol{P}\times\boldsymbol{H}) - \frac{1}{c}\frac{\partial}{\partial t}(\boldsymbol{M}\times\boldsymbol{E}) + \frac{4\pi}{c}\frac{\partial}{\partial t}(\boldsymbol{P}\times\boldsymbol{M})$$

$$= \nabla_{\boldsymbol{E}}(\boldsymbol{P}\cdot\boldsymbol{E}) - \nabla\cdot(\boldsymbol{P}\otimes\boldsymbol{E}) + \nabla_{\boldsymbol{H}}(\boldsymbol{M}\cdot\boldsymbol{H}) - \nabla\cdot(\boldsymbol{M}\otimes\boldsymbol{H}) + \frac{1}{c}\frac{\partial}{\partial t}(\boldsymbol{P}\times\boldsymbol{H} - \boldsymbol{M}\times\boldsymbol{E} + 4\pi\boldsymbol{P}\times\boldsymbol{M})$$

**Part 4.** Identities relating torque to the spin force

$$\partial_j(\boldsymbol{r}\times\hat{\boldsymbol{x}}_i E_i A_j) = (\hat{\boldsymbol{x}}_j\times\hat{\boldsymbol{x}}_i)E_i A_j + \boldsymbol{r}\times\hat{\boldsymbol{x}}_i(\partial_j E_i)A_j + \boldsymbol{r}\times\hat{\boldsymbol{x}}_i E_i(\partial_j A_j)$$
$$= \boldsymbol{A}\times\boldsymbol{E} + \boldsymbol{r}\times\{(\boldsymbol{A}\cdot\nabla)\boldsymbol{E}\} + \boldsymbol{r}\times\{\boldsymbol{E}\cdot(\nabla\cdot\boldsymbol{A})\} = \boldsymbol{A}\times\boldsymbol{E} + \boldsymbol{r}\times\{\nabla\cdot(\boldsymbol{A}\otimes\boldsymbol{E})\}$$

From this $\boldsymbol{P}\times\boldsymbol{E} = \partial_j(\boldsymbol{r}\times\hat{\boldsymbol{x}}_i E_i P_j) - \boldsymbol{r}\times\{\nabla\cdot(\boldsymbol{P}\otimes\boldsymbol{E})\}$ and after integration we get (10a)

**Part 5.** Consider torque on electrons in the field of an SPP

$$\boldsymbol{\tau} = \frac{1}{2}\mathrm{Re}\{\boldsymbol{P}\times\boldsymbol{E}^*\} = \frac{1}{2}\mathrm{Re}\left\{\frac{\chi}{k_0^2|\varepsilon|}(-\hat{\boldsymbol{z}}k_x + \hat{\boldsymbol{x}}i\xi)\times(-\hat{\boldsymbol{z}}k_x - \hat{\boldsymbol{x}}i\xi)\right\}e^{-2\xi z} = \mathrm{Re}\left\{\frac{\chi k_x i\xi}{k_0^2|\varepsilon|}\right\}\hat{\boldsymbol{y}}e^{-2\xi z}$$

$$\bar{\boldsymbol{\tau}} = \frac{1}{2}\mathrm{Re}\left\{\frac{i\chi k_x}{k_0^2|\varepsilon|^2}\right\}\hat{\boldsymbol{y}} = -\frac{\varepsilon''}{8\pi}\frac{k_x}{k_0^2|\varepsilon|^2}\hat{\boldsymbol{y}}$$

Compare it to the energy absorption rate

$$Q = -\frac{\omega}{8\pi}\varepsilon''|\boldsymbol{E}|^2 = \frac{\omega}{8\pi}\frac{\varepsilon''}{k_0^2|\varepsilon|^2}(-\hat{\boldsymbol{z}}k_x + \hat{\boldsymbol{x}}i\xi)(-\hat{\boldsymbol{z}}k_x - \hat{\boldsymbol{x}}i\xi)e^{-2\xi z} = \frac{\omega}{8\pi}\frac{\varepsilon''}{k_0^2|\varepsilon|^2}(k_x^2 + \xi^2)e^{-2\xi z}$$

$$\bar{Q} = \frac{\omega}{8\pi}\frac{\varepsilon''}{k_0^2|\varepsilon|^2}\frac{(k_x^2+\xi^2)}{2\xi}$$

The ratio between them is

$$\frac{\bar{\boldsymbol{\tau}}}{\bar{Q}} = \frac{\frac{\varepsilon''}{8\pi}\frac{k_x}{k_0^2|\varepsilon|^2}}{\frac{\omega}{8\pi}\frac{\varepsilon''}{k_0^2|\varepsilon|^2}\frac{(k_x^2+\xi^2)}{2\xi}} = \frac{\hbar}{\hbar\omega}\frac{2\xi k_x}{(k_x^2+\xi^2)} = \frac{\hbar S_3}{\hbar\omega}$$